\documentclass[lineno]{jfm}
\usepackage{subfigure}
\usepackage{graphicx}
\usepackage{gensymb}
\usepackage{longtable}
\usepackage{float}
\usepackage{epstopdf}
\usepackage{epsfig}
\usepackage[figuresright]{rotating}
\usepackage{adjustbox}
\usepackage{makecell}
\usepackage{bm}
\usepackage{newtxtext}
\usepackage{newtxmath}
\usepackage{caption}
\usepackage{amsmath,mathtools}
\usepackage{natbib}
\usepackage{hyperref}
\hypersetup{
	colorlinks = true,
	urlcolor   = red,
	citecolor  = blue,
}

\newcommand{\RomanNumeralCaps}[1]
\linenumbers

\title{Temperature jump coefficients in polyatomic gas with strong translational/internal non-equilibrium}

\author{Bin Hu \and Lei Wu
	\corresp{\email{wul@sustech.edu.cn}}
}

\affiliation{Department of Mechanics and Aerospace Engineering, Southern University of Science and Technology, Shenzhen 518055, China }

\begin{document}

\maketitle
 
\begin{abstract}
The temperature jump problem in polyatomic gas flow with translational/internal non-equilibrium is investigated, where the internal temperature is excited by volumetric heating, while the translational temperature is heated via the inelastic translational-internal energy relaxation. In the near-continuum flow regime, analytical temperature profiles are derived, based on which the first and second temperature jump coefficients are extracted from the numerical solution of the Boltzmann equation. Analytical expressions for temperature jump coefficients are fitted over a wide range of inelastic collision number and accommodation coefficient.
\end{abstract}		

\section{Introduction} \label{sec:introduction}

The rarefied gas flow is characterized by the Knudsen number (Kn, the ratio of the molecular mean free path $\lambda$ to the characteristic flow length $L_0$), which is described by the Boltzmann equation from the continuum to free-molecular flow regimes. In the case of a small Knudsen number, the gas flow in the bulk is modeled by the Navier–Stokes equation, but rarefaction effects dominate in the Knudsen layer, with a thickness of $O(\lambda)$ adjacent to the solid surface. Due to the complexity of the Boltzmann equation, particularly in engineering problems, the influence of the Knudsen layer is replaced by the boundary conditions of velocity slip and temperature jump. These conditions, in conjunction with the Navier-Stokes equation, offer a pragmatic methodology for depicting the flow dynamics in the near-continuum regime.

Extensive works have been carried out to calculate the temperature jump coefficient (TJC) in monatomic gas flows, see the comprehensive review of~\cite{Sharipov_data2011}.
By matching the kinetic solutions inside the Knudsen layer with the outer Navier-Stokes solution, the first-order TJC is derived from the linearized Boltzmann equation~\citep{Sone1989}, as well as the simplified kinetic models~\citep{BASSANINI1967,SHARIPOV2003, ellipsoidal2009}. The first-order TJC is applied only when $\text{Kn}<0.1$. 
The second TJC, which has been calculated in the steady heat conduction induced by volumetric heating~\citep{Radtke2012second-order} and the unsteady flow induced by time-dependent wall temperature~\citep{Takata2012}, enlarges the application range of the Navier–Stokes equation. 

% An asymptotic expansion of the velocity distribution function has been conducted to extend its applicability, leading to the derivation of the second-order TJC~\citep{sone2007molecular}. 

\begin{figure}
    \centering
    \includegraphics[width=0.5\textwidth]{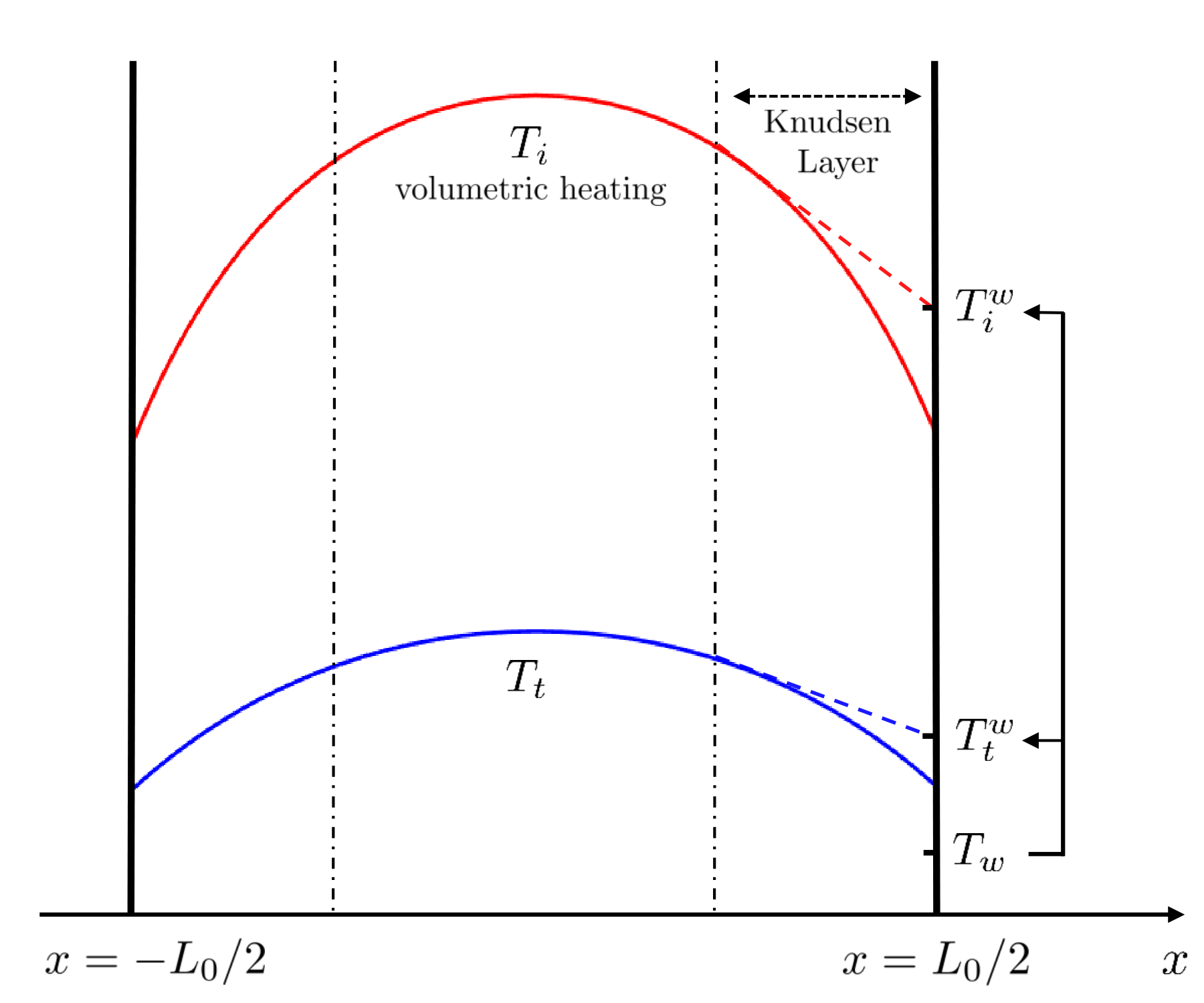}
    \caption{
    Schematic of the temperature profiles of the polyatomic gas between two parallel walls. The internal temperature $T_i$ increases due to volume heating, which subsequently increases the translational temperature $T_t$ through internal-translational energy exchange. The translational and internal temperature jumps are respectively $T_t^w-T_w$ and $T_i^w-T_w$ ($w$ stands for wall).
    }\label{Demo}
\end{figure}

For polyatomic gases, besides the translational temperature, there are internal temperatures linked to rotational, vibrational, or electronic degrees of freedom. The fundamental distinction between polyatomic and monatomic gases is the presence of inelastic translational-internal energy exchange, which significantly impact heat transfer and, consequently, the first and second TJCs. TJCs are derived under conditions where the translational and internal temperatures are identical or closely aligned~\citep{su2022temperature}. However, in high-speed rarefied flows, a divergence between translational and internal temperatures is common~\citep{Hypersonic2019}. For instance, at the exit of a wind tunnel, due to rapid expansion and slow translational-internal energy exchange, the vibrational temperature can be substantially higher than the translational one. The effect of this non-equilibrium on the first and second TJCs has not been previously investigated.
To address this, we consider the polyatomic gas in figure~\ref{Demo}. We shall extract the TJCs by matching the analytical solutions of Navier-Stokes equation with the numerical solutions of the Boltzmann equation. 

% where the internal temperature $T_i$ is increased by volumetric heating, while the translational temperature $T_t$ increases only via the internal-translational energy exchange. Therefore, $T_i$ is always larger than $T_t$, see 

% According to the second-order TJCs, the second TJCs for internal temperature can be obtained from the analytical solution of the temperature profiles. Based on the value of the temperature jump coefficient, it is fitted as a function of the internal collision number and the accommodation coefficient. 

% Surprisingly, the research in this direction is rare. 

% The relaxation rate between the translational and internal energies determines the ratio of the bulk viscosity to the shear viscosity, the thermal relaxation rate of translation and internal heat flux determines the thermal conductivity, including the translational and the internal parts~\citep{wu2020extraction, Li2021JFM}.

% The remaining part of this paper is organized as follows: the kinetic model equation and boundary conditions are described in Section \ref{sec: Kinetic model equation}; the macroscopic equation and the analytical solution of the temperature profile are presented in Section \ref{sec: Flow dynamics in near-continuum regime}; In Section \ref{sec: Extraction of TJCs}, we validate our model against the temperature profiles from the kinetic model and then describe the extraction of TJCs. Finally, the conclusion, as well as future outlooks, are presented in Section \ref{sec: conclusion}.

\section{Kinetic model equation}\label{sec: Kinetic model equation}
Without losing of generality, we consider the polyatomic gas with internal degrees of freedom $d$. Two velocity distribution functions, $f_0(x,\bm v)$ and $f_1(x,\bm v)$ are used to describe the translational and internal states of gas molecules, where $x$ is the spatial coordinate, $\bm v=(v_1,v_2,v_3)$ is the molecules translational velocity. Macroscopic quantities, such as the mass density $\rho$, flow velocity $\bm u$, translational and internal temperatures ($T_t$ amd $T_i$), and translational and internal heat flux ($\bm q_t$ and $\bm q_i$), are obtained by taking moments of distribution functions:
\begin{equation}\label{eq4}
\left[\rho,\rho\bm u,\frac{3}{2}\rho{T_t},\bm q_t\right]= 
\int \left[1,\bm v,\bm c{c^2}\right]f_0d\bm v,
\quad
 \left[\frac{d}{2}\rho  T_i,\bm q_i\right] 
= \int \left[1,\bm c\right]f_1d\bm v,
\end{equation} 
where $\bm c = \bm v -\bm u$ is the peculiar velocity. The total temperature is defined as $T = (3T_t+dT_i)/(3+d)$. The translational and total pressures are $p_t=\rho T_t$ and $p=\rho T$, respectively.

The dynamics of polyatomic gas flow is described by the~\cite{WCU1951} equation, which is even more complicated than the Boltzmann equation. Therefore, the following kinetic equation, which is much simplified but retains the essential physics of energy exchange and the relaxation of translational/internal heat fluxes, is used~\citep{Li2021JFM}:
\begin{equation}\label{kineticEQ}
\begin{aligned}
	& { v_1}\frac{\partial {f}_0}{\partial {x}} = \frac{{f}_{0t}-{f}_0}{{\tau}} + \frac{{f}_{0i}-{f}_{0t}}{Z{\tau}}, \\
	& {v_1} \frac{\partial {f}_1}{\partial {x}} = \frac{{f}_{1t}-{f}_1}{{\tau}} + \frac{{f}_{1i}-{f}_{1t}}{Z{\tau}}
 +\varepsilon\frac{\exp(- \bm c^2)}{\pi^{3/2}}, 
\end{aligned}
    \end{equation}
where ${\tau}$ and $Z\tau$ are the relaxation time of elastic and inelastic collisions, and $Z$ is the internal collision number. The elastic collision conserves the translational energy, while the inelastic collision exchanges the translational and internal energies. $\varepsilon$ is the dimensionless volumetric heating rate, which is similar to the one used in monatomic gas \citep{radtke2011low}. For the flow in figure~\ref{Demo}, we choose $\varepsilon=0.01$, so that the volumetric heating only slightly increases $T_i$, and consequently $T_t$, while the density and velocity remains un-perturbed. Therefore, 
the normalized relaxation time in \eqref{kineticEQ} is $ {\tau}=\sqrt{\frac{4}{\pi}}
   \text{Kn}$.
Finally, the reference distribution functions are given by:
% $f_{0t}(t,x,\bm v)$, $f_{0i}(t,x,\bm v)$, $f_{1t}(t,x,\bm v)$ and $f_{1i}(t,x,\bm v)$ 
 \begin{equation}\label{reference_eq}
		\begin{aligned}
            {f}_{0t} &=\frac{\rho}{(\pi {T}_t)^{-3/2}} \exp\left(\frac{-{\bm c}^2}{{T}_t}\right)\left[1+\frac{4 {\bm q}_t\cdot {\bm c}}{15{T}_t{p}_t}\left(\frac{{\bm c}^2}{{T}_t}-\frac{5}{2}\right)\right], \\
            {f}_{0i} &=\frac{\rho}{(\pi {T})^{3/2} }\exp\left(\frac{-{\bm c}^2}{{T}}\right)\left[1+\frac{4 {\bm q}_0\cdot {\bm c}}{15{T}{p}}\left(\frac{{\bm c^2}}{{T}}-\frac{5}{2}\right)\right], \\ 
            {f}_{1t} &=\frac{d}{2}{T_i}{f}_{0t} + \frac{1}{(\pi {T_t})^{3/2}}
            \exp\left(\frac{{\bm c}^2}{{T_t}}\right)
            \frac{{\bm q}_i \cdot {\bm c}}{{T_t}} ,  \\
            {f}_{1i} &=\frac{d}{2}{T}{f}_{0t} +\frac{1}{(\pi {T})^{3/2}} 
            \exp\left(\frac{{\bm c}^2}{{T}}\right)
            \frac{{\bm q}_1 \cdot {\bm c}}{{T}},   
		\end{aligned}
    \end{equation}
where $\bm q_0$ and $\bm q_1$ are two auxiliary heat fluxes which are defined as the linear combinations of the translational and internal heat fluxes~\citep{Gorji2013,Li2021JFM}:
\begin{equation}\label{eq3}
 \bm q_0=\left(1-\frac{5}{2}\frac{d}{3+d} \right)\bm q_t+\frac{15}{2(3+d)} \bm q_i, \quad 
 \bm q_1=\frac{d}{2(3+d)} \bm q_t+\left[(1-\text{Sc})Z-\frac{3}{2(3+d)}\right] \bm q_i,
\end{equation}
with $\text{Sc}$ being the Schmidt number. %The specific forms are derived from the asymptotic Chapman-Enskog expansion based on the WCU equation.

% In the limit of no translational-internal energy exchange ($Z\rightarrow\infty,A_{tt}=2/3$ and $A_{ti}=A_{it}=A_{ii}=0$), the kinetic model \eqref{kineticEQ} reduces to the \cite{Generalization1968} model equation for monatomic gases.
% When the steady state is reached, the governing kinetic equation can be written in the following dimensionless form as

The spatial coordinate and density is normalized by the reference length $L_0$ and density $\rho_0$, respectively; the temperatures are normalized by the reference temperature $T_w$; the velocity is normalized by the most probable speed $v_0 = \sqrt{2RT_w}$, with $R$ being the specific gas constant; the heat flux is normalized by ${\rho}_0RT_wv_0$. The first and second distribution functions are normalized by $\rho_0/v_0^3$ and $\rho_0RT_w/v_0^3$, respectively.

\section{Analytical temperature profile in the near-continuum regime}\label{sec: Flow dynamics in near-continuum regime}

% In this section, we investigate the macroscopic equation in the near-continuum regime from the kinetic equations. Considering a polyatomic gas confined between two parallel plates, the second-order temperature jump solution is derived by integrating the analytical solution of the temperature profile with the TJC.
% We choose $\varepsilon=0.01$, and the temperature below are the normalized perturbed temperatures $(T-T_w)/T_w$. 
% (normalised by $2k_B\mu(T_w)/m$)

According to the Chapman-Enskog expansion \citep{CE}, the constitutive relations for the heat fluxes can be derived from the first-order expansion of the Knudsen number~\citep{aoki2020two}:
\begin{equation}\label{eq9}
     \begin{aligned}
        \bm q_t\approx \bm q^{NS}_t = -\kappa_t\nabla T_t,\quad
        \bm q_i\approx \bm q^{NS}_i = -\kappa_i\nabla T_i,
     \end{aligned}
    \end{equation}
where $\kappa_t$ and $\kappa_i$ are the dimensionless transitional and internal thermal conductivities~\citep{Li2021JFM,Su2021CMAME}:
\begin{equation}{\label{Eucker_factor}}
    \begin{aligned}[b]
        \kappa_t =\frac{15\tau}{8}\left[1-\frac{5d}{4(d+3)Z}\left(1-\frac{2}{5\text{Sc}}\right)\right], \quad
        \kappa_i = \frac{d\tau}{4\text{Sc}}\left[1+\frac{15}{4(d+3)Z}\left(1-\frac{2}{5\text{Sc}}\right)\right].
    \end{aligned}
\end{equation}
Therefore, the total and internal energy conservation equations can be written as (Only the energy equations need to be considered in the problem in figure~\ref{Demo}):
\begin{equation}\label{temperature_governing}
     \begin{aligned}
       & {\kappa}_t\frac{\partial^2 {T}_t}{\partial {x}^2} +{\kappa}_i\frac{\partial^2 {T}_i}{\partial {x}^2} = -\varepsilon,\\
      &  {\kappa}_i\frac{\partial^2 {T}_i}{\partial {x}^2} = - \varepsilon + \frac{3d}{2Z{\tau}(3+d)}({T}_i-{T}_t).
     \end{aligned}
\end{equation}

By considering the symmetry condition at $x=0$, the second order ordinary differential equation system \eqref{temperature_governing} for the two temperatures $T_i$ and $T_t$ can be solved as follows:
\begin{equation}\label{temperature_solution}
    \begin{aligned}
        &{\Delta T}_i  = C_2\left[\exp(A{x})+\exp(-A{x})\right] + C{x}^2 + \frac{C_1}{\kappa_t + \kappa_i}+\frac{\kappa_t}{(\kappa_t + \kappa_i)\kappa_iA^2},    \\ 
         &{\Delta T}_t  = -\frac{\kappa_i C_2}{\kappa_t}\left[\exp(A{x})+\exp(-A{x})\right] + C{x}^2 + \frac{C_1}{\kappa_t + \kappa_i}-\frac{1}{(\kappa_t + \kappa_i)A^2}, \\ 
        &{\Delta T}= \frac{(d\kappa_t - 3\kappa_i)C_2}{(3 + d)\kappa_t} \left[ \exp(Ax) + \exp(-Ax) \right] + Cx^2 + \frac{C_1}{\kappa_t + \kappa_i} + \frac{d\kappa_t - 3\kappa_i}{\kappa_i(3 + d)(\kappa_t + \kappa_i)A^2},
\end{aligned}   
\end{equation}
where ${\Delta T}_{i} = (T_{i}-T_w)/\varepsilon$, ${\Delta T}_{r} = (T_{r}-T_w)/\varepsilon$, $\Delta T = (T-T_w)/\varepsilon$, $A^2=3d({\kappa}_t+{\kappa}_i)/2Z{\tau}{\kappa}_t{\kappa}_i(3+d)$, $C=-{1}/{2({\kappa}_t+{\kappa}_i)}$, and the two free parameters $C_1$, $C_2$ need to be determined by the boundary conditions. 

According to the work of~\cite{Takata2012}, the jump boundary conditions for the translational and internal temperatures can be heuristically written as:
\begin{equation}\label{ord2}
\begin{aligned}
            T_i-{T}_{w} = d_1^i{\tau}\frac{\partial {T_i}}{\partial {x}}\bigg|_{w}
            +d_2^i{{\tau}}^2\frac{\partial^2 {T_i}}{\partial {x}^2}\bigg|_{w}, \quad
            T_t-{T}_{w} = d_1^t{\tau}\frac{\partial {T_t}}{\partial {x}}\bigg|_{w}
            +d_2^t{{\tau}}^2\frac{\partial^2 {T_t}}{\partial {x}^2}\bigg|_{w},
\end{aligned}   
\end{equation}
where $d_1^{i}$ and $d_2^{i}$ are the first and second TJCs for the internal temperature, while $d_1^t$ and $d_2^t$ are the first and second TJCs for the translational temperature, respectively. 
Combining \eqref{temperature_solution} and \eqref{ord2}, the analytical second-order temperature jump solution (STJ) for the internal and translational temperature can be obtained:
\begin{equation}\label{analytical}
\begin{aligned}
\Delta T_i &= C_2 \Bigg\{ \exp(Ax) + \exp(-Ax) + d_1^{i}\tau A \left[ \exp\left(-\frac{A}{2}\right) - \exp\left(\frac{A}{2}\right) \right] \\
&+ (d_2^{i}\tau^2 A^2 - 1) \left[ \exp\left(-\frac{A}{2}\right) + \exp\left(\frac{A}{2}\right) \right] \Bigg \} + C\left( x^2 - \frac{1}{4} - d_1^{i}\tau + 2d_2^{i}\tau^2 \right), \\
\Delta T_t &= -\frac{\kappa_i C_2}{\kappa_t} \Bigg\{ \exp(Ax) + \exp(-Ax) + d_1^t\tau A \left[ \exp\left(-\frac{A}{2}\right) - \exp\left(\frac{A}{2}\right) \right] \\
& + (d_2^t\tau^2 A^2 - 1) \left[ \exp\left(-\frac{A}{2}\right) + \exp\left(\frac{A}{2}\right) \right] \Bigg \} + C\left( x^2 - \frac{1}{4} - d_1^t\tau + 2d_2^t\tau^2 \right),
\end{aligned}
\end{equation}
where
the parameter $C_2$ is determined by aligning the internal temperature in the STJ model with the kinetic solution at $x=0$:
\begin{equation}\label{analytical_c21}
    \begin{aligned}
       C_2 &= \frac{{\Delta T_i}\big|_{x=0}+C\left(\frac{1}{4}+d_1^{i}{\tau}-2d_2^{i}{{\tau}}^2\right)}{\left \{2+d_1^{i}{\tau}A\left[\exp\left(-\frac{A}{2}\right)-\exp\left(\frac{A}{2}\right)\right]+(d_2^{i}{{\tau}}^2A^2-1)\left[\exp\left(-\frac{A}{2}\right)+\exp\left(\frac{A}{2}\right)\right]\right \}},
    \end{aligned}
\end{equation}
and by aligning the translational temperature in the STJ model with the kinetic solution at $x=0$:
\begin{equation}\label{analytical_c22}
    \begin{aligned}
       C_2 = -\frac{\kappa_t}{\kappa _i} \frac{{\Delta T_t\big|_{x=0}}+C\left(\frac{1}{4}+d_1^t\tau-2d_2^t\tau^2\right )}{\left \{2+d_1^t{\tau}A\left[\exp\left(-\frac{A}{2}\right)-\exp\left(\frac{A}{2}\right)\right]+(d_2^t{{\tau}}^2A^2-1)\left[\exp\left(-\frac{A}{2}\right)+\exp\left(\frac{A}{2}\right)\right]\right \} }.
    \end{aligned}
\end{equation}
All temperatures can be determined via \eqref{analytical} after $C_2$ is determined by a temperature jump boundary condition. Similarly, the parameter $C_1 = (\kappa_t\Delta T_t + \kappa_i\Delta T_i)|_{x=0}$ is determined to guarantee the translational or internal temperature not determined by the parameter $C_2$ in the STJ model agree with the kinetic solution at $x=0$.

\section{Extraction of temperature jump coefficients}\label{sec:Maxwell_TJCs}

In this section, the first and second TJCs are extracted by matching the analytical solutions~\eqref{analytical} with the numerical solutions of the kinetic equation~\eqref{kineticEQ}, which are obtained by the discrete velocity method~\citep{Su2021CMAME}.

\subsection{Maxwell's specular-diffuse boundary condition}

\begin{figure}
    \centering
    \includegraphics[width=0.85\textwidth]{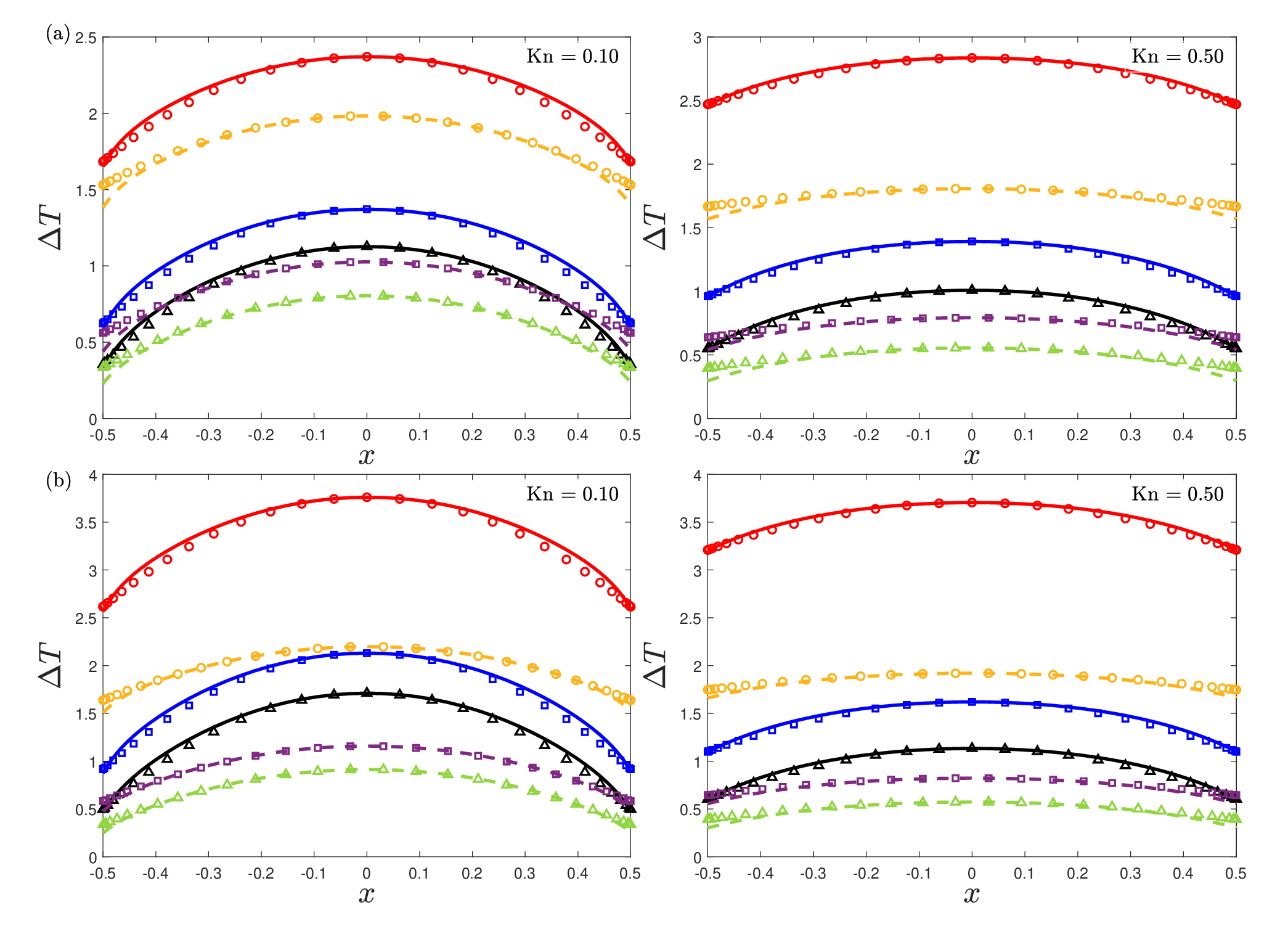}
    \caption{
    Comparisons between the STJ model and the kinetic model for \eqref{TJCs} or \eqref{TJCs_t} when $Z=10$  (a) and 100 (b). Solid ($T_i$) and dashed ($T$) lines represent the solution of the kinetic equation~\eqref{kineticEQ}, while circles, squares, and triangles are the STJ solutions \eqref{analytical} with $\alpha_0= $0.3, 0.7, and 1, respectively.}
    \label{fig:Maxwell}
\end{figure}

We first consider the Maxwellian specular-diffuse boundary condition, where a fraction $\alpha_0$ of gas molecules striking the solid wall are diffusely reflected, and the remainder are specularly reflected. 
Figure~\ref{fig:Maxwell} shows the typical temperature profiles from the kinetic equation, when $\text{Kn}=0.1$ and 0.5, and when the inelastic collision number $Z=10$ and 100. It is seen that, first,
increasing Kn diminishes the curvature of the temperature profiles, but increases the temperature jump at the solid wall. 
Second, increasing $Z$ induces a distinct segregation of the internal and total temperature profiles. This is because, when $Z$ increases, the internal-translational energy relaxation slows, and the transfer from the internal energy to translational energy is reduced. Note that when Z $\rightarrow 1000$, there is almost no translational-internal energy transfer, so that the translational temperature is barely heated. 
Third, an increase in the accommodation coefficient $\alpha_0$ indicates a more significant proportion of energy being diffusely reflected, hence increasing the energy absorption by the solid wall. As a consequence,  both translational and internal temperatures decrease. 

%As for the STJ model, the slope of the temperature profiles is primarily determined by the value of parameter $A$. Through approximate simplification for the parameter $A^2=\pi(3f_{ti}+df_{it})/2(3+d)f_{ti}f_{it}Z\text{Kn}^2$, it is found that the proportional relationship between $A$ and $1/\text{Kn}$ which explains why an increase in Kn reduces the curvature of the temperature curve. As the accommodation coefficient $\alpha_0$ decreases, the TJCs increase to capture the more intense boundary temperature jump phenomenon.

% Within our optimization framework, considering the second-order TJCs within the Knudsen number Kn extending from 0.05 to 0.5, alongside the internal collision number $Z$ that spans between 1 and 10000 and the accommodation coefficient $\alpha_0$ ranges from 0.0001 to 1. Under our parameter configuration of $Z$,$\text{Kn}$,$\alpha_0$, the transitional temperature approaches zero in most cases. In order to more effectively reflect the performance of the TJCs, we opt for the internal energy temperature and total temperature profiles. The objective is to minimize the difference between the STJ and the kinetic models concerning the internal and total temperature profiles.

We extract the TJCs over a range of Knudsen numbers Kn from 0.05 to 0.5, the internal collision number $Z$ from 1 to 10,000, and the accommodation coefficient $\alpha_0$ from 0.0001 to 1. Therefore, by adjusting the TJCs $d_1^{i}$ and $d_2^{i}$, we minimize the following quantity: 
\begin{equation}\label{opt}
\begin{aligned}
       L(d_1^{i},d_2^{i}) = \sum^{Z}\sum^{\alpha_0}\sum^{\text{Kn}}
       \int\left(|{T_t}^{\text{STJ}}-{T}_t^{\text{K}}|+|{T}_i^{\text{STJ}}-{T}_i^{\text{K}}|\right)dx,
\end{aligned}   
\end{equation}
which is the sum of the integral of the absolute difference between the STJ and the kinetic model in the whole $x$ coordinate interval.
The superscript $\text{K}$ and $\text{STJ}$ denotes temperatures from the kinetic and STJ models, respectively. 

With the fixed number of the internal collision number $Z$ and the accommodation coefficient $\alpha_0$, the TJCs remain invariant when the Knudsen number is small. By minimizing the sum of the Knudsen number in the quantity~\eqref{opt}, we find the TJCs with the smallest temperature difference between the kinetic model and the STJ model for all ranges of Knudsen numbers. Our investigation reveals that when the internal collision number $Z$ and the accommodation coefficient $\alpha_0$ vary, the TJCs change with $\alpha_0$ according to the scaling law $(2-\alpha_0)/\alpha_0$. Also, they increase with $Z$ at a decreasing rate, eventually saturate at higher values of $Z$. To capture this behavior, we use the Padé approximation $(aZ+b)/(Z+c)$ with the condition of $b-ac<0$ and the scaling law $(2-\alpha_0)/\alpha_0$ to model and fit the TJCs as a function of $Z$ and $\alpha_0$. The particle swarm optimization algorithm~\citep{2011Particle} is employed to minimize the quantity \eqref{opt} with $C_2$ given by \eqref{analytical_c21}, resulting in the following formula:
\begin{equation}\label{TJCs}
    \begin{aligned}
    & d_1^{i}=\frac{2-\alpha_0}{\alpha_0} \frac{1.013 Z-0.760}{Z-0.726}, \quad
 d_2^{i}=\frac{2-\alpha_0}{\alpha_0} \frac{0.452 Z-0.024}{Z+0.695}.
    \end{aligned}
\end{equation}

%  It is proposed that TJCs vary with $\alpha_0$ according to a $(2-\alpha_0)/\alpha_0$ scaling law. Additionally, our investigation reveals that TJCs increase with $Z$ at a decreasing rate, eventually saturate at higher values of $Z$. To capture this behavior, we employ the Padé approximation $(aZ+b)/(Z+c)$ with the condition of $b-ac<0$ to fit the optimal TJCs effectively. We optimize the temperature difference between the kinetic model and STJ \eqref{analytical} for all the Knudsen numbers under the given internal collision number Z. Ultimately, the TJCs are modeled as functions of $Z$ and $\alpha_0$. This relationship is encapsulated in the following mathematical formulation by using the Particle Swarm Optimization~\citep{2011Particle}: 

%With \eqref{analytical_c2} and \eqref{TJCs}, we can obtain the translational TJCs $d_1^{t},d_2^{t}$ numerically. 

Similarly, when the particle swarm optimization algorithm~\citep{2011Particle} is employed to minimize the quantity \eqref{opt} with $C_2$ given by \eqref{analytical_c22}, resulting in the following formula:
%we provide a fitted form of $d_1^{t},d_2^{t}$ for practical use. By replacing $d_1^{i},d_2^{i}$ by $d_1^{t},d_2^{t}$ in quantity \eqref{opt}, the translational TJCs can be obtained as:
\begin{equation}\label{TJCs_t}
    \begin{aligned}
    & d_1^{t}=\frac{2-\alpha_0}{\alpha_0} \frac{9.791 Z-4.432}{Z-8.820}, \quad
 d_2^{t}=\frac{2-\alpha_0}{\alpha_0} \frac{9.070 Z +2.346}{Z+9.254}.
 % 9.790954835	-4.431562139	8.819503679	9.069723515	2.345573599	9.253979912
    \end{aligned}
\end{equation}
The comparative results of temperature profiles between the kinetic model and the STJ approach are presented in figure~\ref{fig:Maxwell}. Here, we notice that the transitional temperature is significantly lower than the internal temperature and the total temperature under our parameter setting. Therefore we only show the internal and total temperature profiles obtained by the internal TJCs, as the performance of the two TJCs models is similar.
Notably, the internal and total temperatures computed by the STJ method closely match those predicted by the kinetic model. This agreement underscores the effectiveness of the TJCs in bridging the STJ model with kinetic simulations, highlighting its utility in accurately simulating temperature distributions across different values of $Z$ and $\alpha_0$.

% \leir{calculate the TJCs for the translation temperature, and give the analytical formula like (4.2).It is difficult to give an explicit expression, which can only be calculated according to $d_1^i$ and $d_2^i$ and the formula (3.6). Under the formula 3.6, how to solve $d_1^t$ and $d_2^t$ is described}

% Furthermore, for a given set of translational/internal accommodation coefficients, it is hypothesized that the TJCs vary with $\alpha_0$ according to a $(2-\alpha_0)/\alpha_0$ scaling law.Meanwhile, concerning the variation of the TJCs with respect to the internal collision number $Z$, our investigation reveals that they increase with $Z$ at a decelerating rate and remain unchanged at a large value of $Z$.The algorithmic approach focuses on optimizing the TJCs across all Knudsen numbers for a specified internal collision number $Z$. Our objective is to identify the optimal TJCs for each specific internal collision number. Subsequently, we investigate the variation of TJCs under different accommodation coefficients $\alpha_0$, scaling them by the ratio $(2-\alpha_0) / \alpha_0$. This modeling process aims to capture their dependency through a rigorous fitting procedure. Specifically, the functional forms employed are designed to comprehensively describe the relationship between TJCs and the variables $Z$ and  $\alpha_0$, ensuring a robust representation suitable for analytical insights and computational implementation. 

\subsection{Different translational/internal energy accommodation coefficients}
In hypersonic flows, the energy accommodation coefficients for the translational and internal modes might be different. For example, in the experiment of hypersonic flows passing over a double cone, the vibrational energy follows a near-specular reflection in order to match the numerical and experimental surface heat flux~\citep{Hypersonic2019,liu2024shock}. 
To this end, we apply the diffuse boundary condition for the translational velocity distribution function, while the specular-diffuse boundary condition for the internal velocity distribution function, i.e., for $f_1$, a fraction $\alpha_i$ of gas molecules striking the solid wall are diffusely reflected, and the remainder are specularly reflected. %A similar boundary condition has been used in explaining the supersonic flows passing over double cone~\citep{Hypersonic2019}.

\begin{figure}
    \centering
    \includegraphics[width=0.85\textwidth]{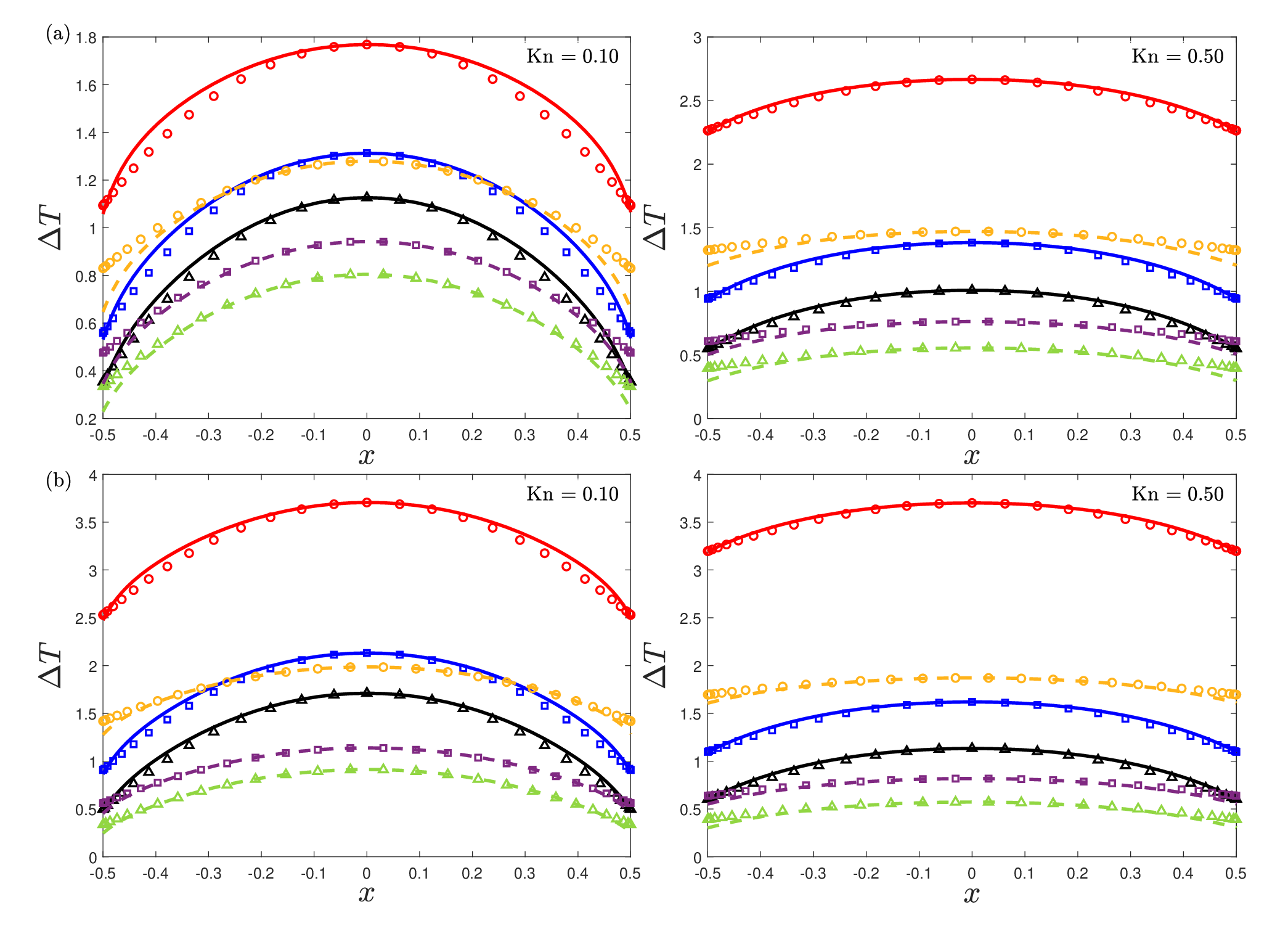}
    \caption{Comparisons between the STJ model and the kinetic model for \eqref{TJCs21} or \eqref{TJCs22} when $Z=10$ (a) and 100 (b). Solid ($T_i$) and dashed ($T$) lines represent the solution of the kinetic model~\eqref{kineticEQ}, while circles, squares, and triangles are the STJ solutions \eqref{analytical} with $\alpha_i=$ 0.3, 0.7, and 1, respectively.}
    \label{fig:Maxwell_different}
\end{figure}

Figure~\ref{fig:Maxwell_different} shows the typical temperature profiles from the kinetic equation \eqref{kineticEQ}, when $\text{Kn}=0.1$ and 0.5, and when the inelastic collision number $Z=10$ and 100. The roles of $\text{Kn}$ and $Z$ in the temperature profile are the same as that in the previous subsection. When $Z$ is large, $\alpha_i$ and Kn change similarly. 
However, when $Z$ is relatively small, the translational temperature is set under diffuse boundary conditions, leading to increase energy absorption by the wall. This results in a reduction of the translational temperature and the total flow energy, which in turn causes a decrease in the internal energy temperature and the total temperature values, thereby increasing their gap. For the STJ model, different boundary conditions setting reduce the temperature ${\Delta T_i}\big|_{x=0}$ along with the minus parameter $C_2$ in \eqref{analytical} compared with the same boundary conditions, thereby amplifying the slope of the temperature profiles.
% However, when $Z$ is small, it is not equal to 1 at $\alpha_i$, because the trFor the STJ model, different boundary conditions setting reduce the temperature ${\Delta T_i}\big|_{x=0}$ along with the minus parameter $C_2$ in \eqref{analytical} compared with the same boundary conditions, thereby amplifying the slope of the temperature profiles.anslational temperature boundary is completely diffuse, the energy absorbed by the wall becomes more, the translational temperature decreases, and the translation-internal energy exchange decreases, the internal energy temperature and the total temperature also decrease and the interval becomes larger. It is notable that maintaining the same parameter configurations, while enhancing wall surface absorption through complete diffuse reflection leads to a decrease in temperatures.
%The variation in $\alpha_i$ under different boundary conditions induce changes in temperature profiles consistent with those observations under the same boundary conditions. 

For different translational/internal accommodation coefficients, the optimization process yields distinct optimal TJCs dependent on both $Z$ and $\alpha_i$. We similarly hypothesize that TJCs vary with Z according to the Padé approximation form, yet their dependence on $\alpha_i$ not only adheres to the scaling law but also involves the linearly fitted parameters $a$, $b$ and $c$.
We also note that in our model, when the accommodation coefficient equals 1, both the transitional temperature and the internal temperature distribution are diffuse boundary conditions, thus the TJCs should reduce to these in \eqref{TJCs} and \eqref{TJCs_t}.
%The boundary condition settings are indistinguishable from   Maxwell’s boundary condition and the TJCs obtained by both are the same in this case. 
%Therefore, the TJCs calculated by \eqref{TJCs} is equivalent to the TJCs of different translational/internal accommodation coefficients when $\alpha_i=1$, and 
Eventually, the TJCs expression obtained by the minimization quantity \eqref{opt} is as follows:
% We also note that, In our model, when the accommodation coefficient equals 1, the reflection model exhibits fully diffuse reflection for both transitional and internal temperature, thereby fitting identical temperature data in both models. We integrate this constraint into the optimization framework to ensure that the TJCs obtained from different translational/internal energy accommodation coefficients setting are the same as Maxwell’s boundary condition. Therefore, we add the temperature distribution differences calculated according to formula \eqref{TJCs} under the same boundary conditions for all the accommodation coefficients in \eqref{opt}. This is equivalent to optimizing the temperature difference simultaneously with TJCs set under two boundary conditions setting. The resulting expression for the TJCs is as follows:
\begin{equation}\label{TJCs21}
    \begin{aligned}
     d_1^{i}=\frac{2-\alpha_i}{\alpha_i} \frac{\bm a_1 Z+\bm a_2}{Z+\bm a_3},\quad d_2^{i}=\frac{2-\alpha_i}{\alpha_i} \frac{\bm b_1 Z+\bm b_2}{Z+\bm b_3},
     \quad
     d_1^{t}=\frac{\bm c_1 Z+\bm c_2}{Z+\bm c_3},\quad 
    d_2^{t}=\frac{\bm d_1 Z+\bm d_2}{Z+\bm d_3},
\end{aligned}
\end{equation}
where the parameters $P^i=[\bm a_1,\bm a_2,\bm a_3,\bm b_1,\bm b_2,\bm b_3]$ and $P^t=[\bm c_1,\bm c_2,\bm c_3,\bm d_1,\bm d_2,\bm d_3]$ within the expression are computed by:
\begin{equation*}\label{coef_alpha_0-1}
    % \begin{bmatrix}
    % \bm a_1\\
    % \bm a_2\\
    % \bm a_3\\
    % \bm b_1\\
    % \bm b_2\\
    % \bm b_3
    % \end{bmatrix}
    P^i
    =
    \begin{bmatrix}
    0.095 & 0.917 & 0.002 \\
    -1.562 & 1.202 & -0.374 \\
    -0.695 & -1.508 & 1.504 \\
    0.082 & 0.368 & 0.002 \\
    0.845 & -0.744 & -0.133 \\
    4.510 & -7.070 & 3.250
    \end{bmatrix}
    \begin{bmatrix}
	\alpha_i\\
	1\\
        {\alpha_i}^{-1}
    \end{bmatrix},
    \quad
    % \begin{bmatrix}
    % \bm c_1\\
    % \bm c_2\\
    % \bm c_3\\
    % \bm d_1\\
    % \bm d_2\\
    % \bm d_3
    % \end{bmatrix}
    P^t
    =
    \begin{bmatrix}
        -0.963 & 11.245 & -0.491 \\
        1.308 & -7.451 & 1.711 \\
        15.484 & -8.633 & 1.968 \\
        -0.853 & 10.358 & -0.435 \\
        0.387 & -1.247 & 3.206 \\
        14.701 & -7.125 & 1.678
    \end{bmatrix}
    \begin{bmatrix}
        \alpha_i\\
        1\\
        \alpha_i^{-1}
    \end{bmatrix}.
\end{equation*}

Figure~\ref{fig:Maxwell_different} presents the comparison of the temperature profiles between the kinetic model and the STJ model results under different boundary conditions. The total and internal temperatures, as determined through the STJ model, exhibit a remarkable alignment with the predictions of the kinetic model. Except that when $Z$, Kn, $\alpha_i$ is comparatively small, the temperature jump phenomenon of the total temperature is severe and the STJ results deviate significantly from the actual boundary temperature values, revealing the limitations of the STJ model in capturing the temperature jump behavior in these specific situation. This observation highlights the importance of carefully considering the selection of boundary conditions and model parameters in the accurate prediction of temperature profiles. 

% For the different translational/internal energy accommodation coefficients, we consider varying boundary conditions which use the fully diffuse reflection for transitional temperature and the diffuse-specular reflection for internal temperature. 
% The reflection kernels for two temperatures are defined as follows:
% \begin{equation}\label{dsm2}
%     \begin{aligned}
%       \mathscr{R}_{T_i}(\bm v'\rightarrow\bm v) &= \alpha_i\frac{|\bm v\cdot\bm n|}{{(T^w_i)}^2}\exp\left(-\frac{\bm v^2}{T^w_i}\right)+(1-\alpha_i)\delta(\bm v'-\bm v+2(\bm v\cdot\bm n)\bm n), \\ 
%       \mathscr{R}_{T_t}(\bm v'\rightarrow\bm v) &= \frac{|\bm v\cdot\bm n|}{{(T^w_t)}^2}\exp\left(-\frac{\bm v^2}{T^w_t}\right). \\ 
%     \end{aligned}
% \end{equation}
% This approach constructs a framework where TJCs are systematically related to $\alpha_i$, ensuring a comprehensive understanding and effective parameterization in our analytical models. Additionally, this approach allows us to compute TJCs with the accommodation coefficient of 1 under same boundary conditions based on parameters from different boundary conditions. Utilizing the scaling law $(2-\alpha_i)/\alpha_i$, we calculate TJCs for all accommodation coefficients $\alpha_i$ and the internal collision numbers $Z$ of the two boundary settings. 

\begin{figure}
	\centering
	\includegraphics[width=0.45\linewidth]{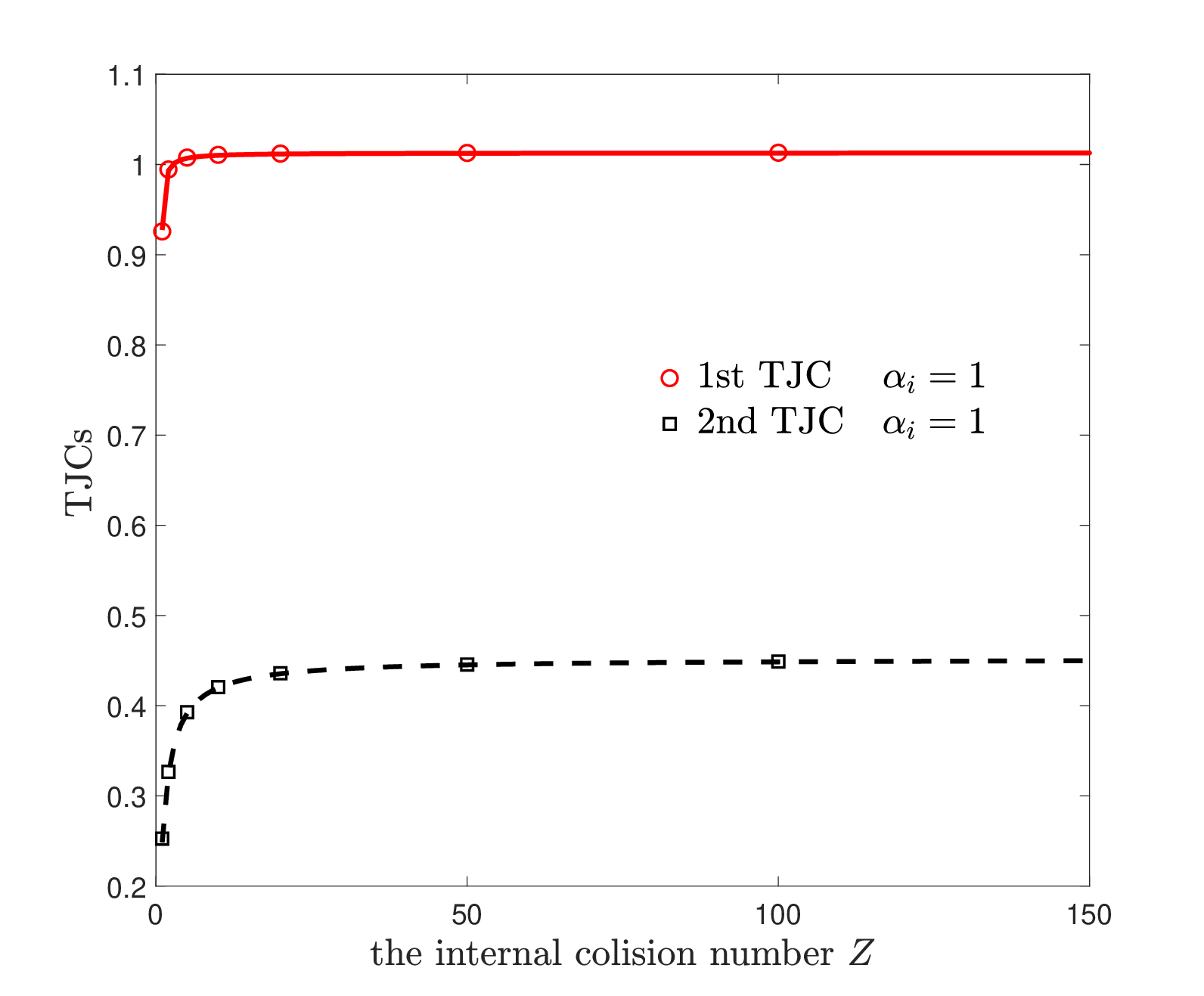}
	\includegraphics[width=0.45\linewidth]{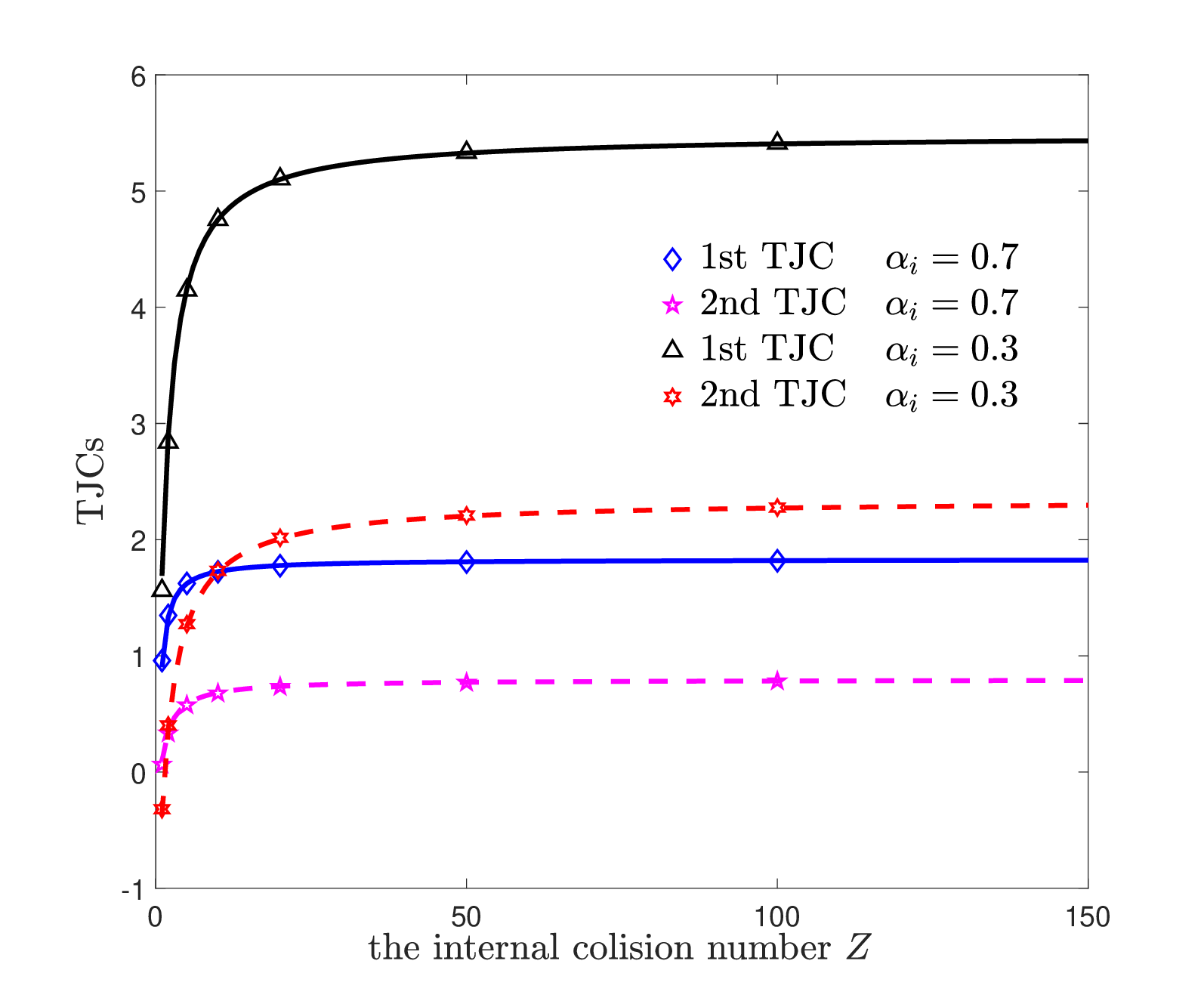}
 \caption{
The TJCs from the fit expression (lines, given by \eqref{TJCs21}) and the optimized data (markers) under different boundary conditions.}\label{fig_tjc}
\end{figure}

Figure~\ref{fig_tjc} shows the comparison of \eqref{TJCs21}  with TJCs of optimized data. It is observed that at elevated Z values, the progression of TJCs is notably slow. Additionally, the first-order TJCs are generally greater second-order TJCs. With a reduction in the accommodation coefficient, the temperature rises, resulting in more pronounced temperature jump phenomena. As a result, the values of TJCs increase and the Z value at which the TJCs shift from an increasing trend to a plateau also augments.

\subsection{The limit of vanishing internal energy accommodation}

In the previous discussion, the computation conducted by TJCs for the internal temperature was based on the conventional range of accommodation coefficients confined between 0.1 and 1 under the condition of fully diffuse reflection for translational temperature. However, our current investigation pivots shifts toward more extreme scenarios where the accommodation coefficient approaches magnitudes around $10^{-3}$, signifying an adoption of a nearly completely specular reflection model for the internal temperature. 

The purpose of using these diverse boundary conditions in our design is to simulate the physical situation of strong transitional/internal energy non-equilibrium. A crucial aspect requiring statement is that under such strong non-equilibrium conditions, the probability of inelastic collisions decreases, resulting in reduced exchange between translational and internal energy. In light of these phenomena, we re-calibrated the optimization bounds for internal collision numbers $Z$ spanning from 50 to 10000. This adjustment is essential for accurately simulating and capturing the temperature jump behavior under strong non-equilibrium conditions observed in actual physical scenarios. In alignment with different translational/internal energy accommodation coefficients, we explored various gas-interface models and engaged in optimization to derive the expressions for the TJCs below:
% Consequently, this causes an increase in internal temperature and a decrease in translational temperature. By fine-tuning the range of internal collision numbers, we enhance the sensitivity of the STJ model to the subtleties of energy exchange processes, thereby providing a robust framework for analyzing and predicting temperature profiles in highly non-equilibrium environments. 
\begin{equation}\label{TJCs3}
    \begin{aligned}
    & d_1^{i}=\frac{2-\alpha_i}{\alpha_i} \frac{\bm a_1 Z+\bm a_2}{Z+\bm a_3}, \quad
     d_2^{i}=\frac{2-\alpha_i}{\alpha_i} \frac{\bm b_1 Z+\bm b_2}{Z+\bm b_3},
     \quad
      d_1^{t}=\frac{\bm c_1 Z+\bm c_2}{Z+\bm c_3},\quad 
      d_2^{t}=\frac{\bm d_1 Z+\bm d_2}{Z+\bm d_3},
\end{aligned}
\end{equation}
where the parameters within the expression are governed by:
\begin{equation*}\label{coef_alpha_small}
 %    \begin{bmatrix}
 %        \bm a_1\\
	% \bm a_2\\
 %        \bm a_3\\
	% \bm b_1\\
	% \bm b_2\\
 %        \bm b_3\\          
 %    \end{bmatrix}
 P^i
    =
    \begin{bmatrix}
    0.998 & 0.973 & 0 \\
    0.983 & -0.527 & 0 \\
    2.257 & 115.289 & 0.936 \\
    0.993 & 0.400 & 0 \\
    0.874 & -10.416 & 0 \\
    1.463 & 43.085 & 2.768
\end{bmatrix}
    \begin{bmatrix}
	\alpha_i\\
	1\\
        {\alpha_i}^{-1}
\end{bmatrix},
\quad
%     \begin{bmatrix}
%     \bm c_1\\
%     \bm c_2\\
%     \bm c_3\\
%     \bm d_1\\
%     \bm d_2\\
%     \bm d_3
% \end{bmatrix}
P^t
=
\begin{bmatrix}
    1.078 & 8.088 & -0.002 \\
    -1.71 & -245.342 & 0.103 \\
    3.112 & 193.036 & -0.033 \\
    0.989 & 0.009 & 0 \\
    -31.912 & -2996.535 & 0.013 \\
    1.809 & 74.538 & 0.009
\end{bmatrix}
\begin{bmatrix}
    \alpha_i\\
    1\\
    \alpha_i^{-1}
\end{bmatrix}.
\end{equation*}

\begin{figure}
    \centering
    \includegraphics[trim={0 20 0 0},clip,width=0.85\textwidth]{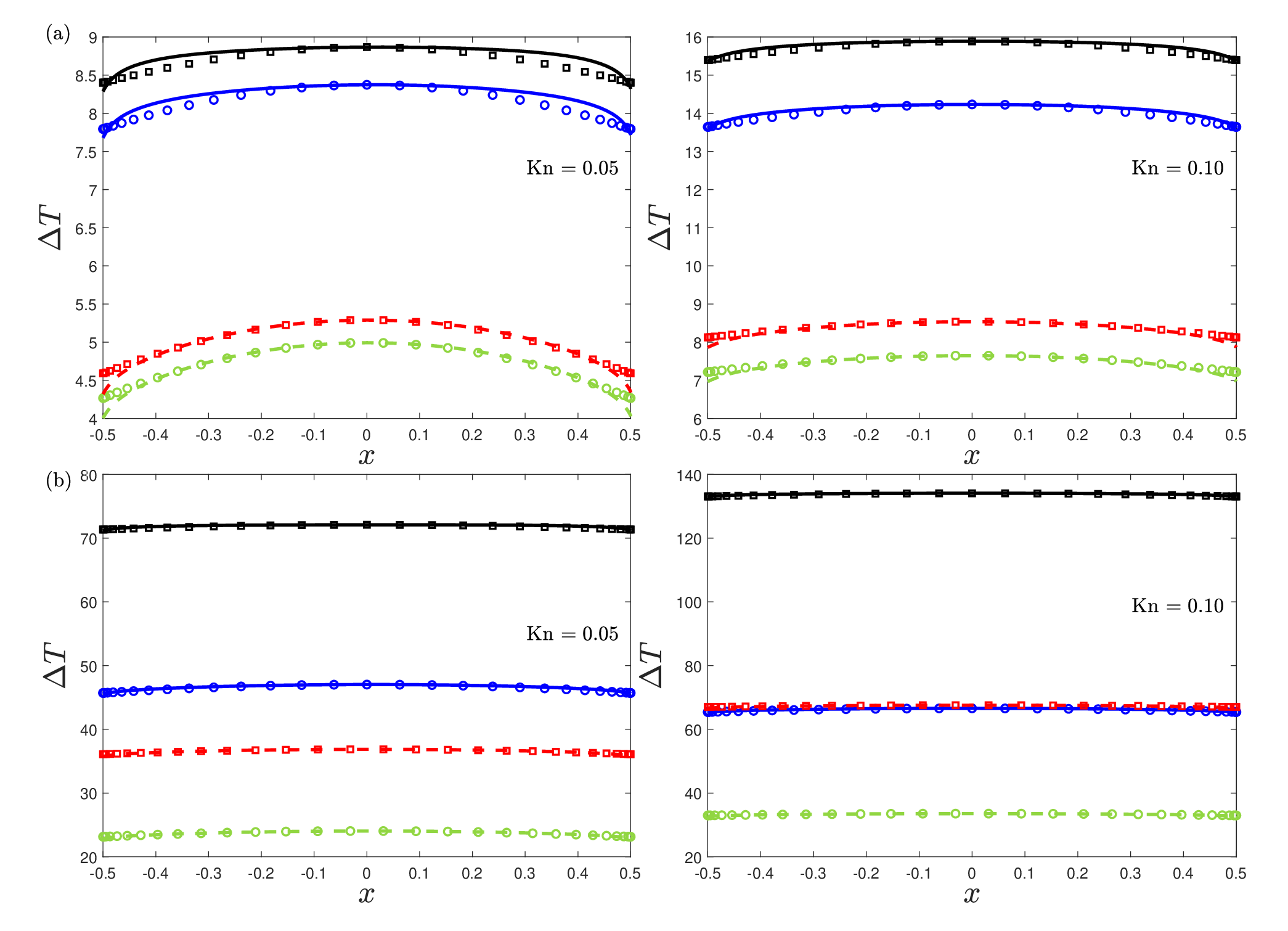} 
    \caption{
    Comparisons between the STJ model and the kinetic model for \eqref{TJCs3} when $Z=100$ and 1000 (b). Solid ($T_i$) and dashed ($T$) lines represent the solution of the kinetic model~\eqref{kineticEQ}, while circles and squares are the STJ solutions \eqref{analytical} with $\alpha_i= 0.01$ and 0.001, respectively.}
    \label{fig4}
\end{figure}

Figure~\ref{fig4} presents a comparative analysis of temperature profiles obtained from the kinetic and STJ models for small internal energy accommodation coefficient values. The internal and total temperatures calculated from the STJ model are in good agreement with those from the kinetic model. As the accommodation coefficient diminishes, the temperature becomes extremely high. 
Furthermore, as $Z$ exceeds 100, the curvature of the temperature profiles plot diminishes, and the phenomenon of internal energy temperature jumps becomes overlooked.
% To better observe examine the characteristics and patterns of the temperature profiles, a logarithmic temperature scale has been applied to certain high-temperature curves.
% The figure reveals that when $Z$ approaches 50, the internal and total temperatures are nearly indistinguishable, with changes induced by the accommodation coefficient being negligible.
% Our objective is to develop a comprehensive set of equations capable of accurately computing TJCs for the entire spectrum of accommodation coefficients.
In the pursuit of enhancing the predictive accuracy of the TJC model, it is crucial to address the limitations observed in the TJCs derived from \eqref{TJCs3}. While these TJCs exhibit satisfactory temperature fitting performance under small value of accommodation coefficients, they fail to yield identical TJCs values obtained from {\eqref{TJCs21}  across the range of accommodation coefficients from 0.1 to 1, as discussed previously. To overcome this challenge, we can undertake an effort to incorporate equations with a higher linearity in their parameter coefficients, thereby ensuring robust fitting performance across the entire spectrum of accommodation coefficients.
%This strategic approach aims to refine the parameters within the TJCs expressions , encompassing both the 0.1 to 1 and $10^{-4}$ to $10^{-2}$ ranges. Nevertheless, this broadening of the parameter space introduces increased sensitivity to the coefficients themselves. The condition number associated with the matrix in \eqref{coef_final} is notably large, indicating that the matrix is ill-conditioned. This ill-conditioning necessitates an exceptional level of precision in our parameter estimation, demanding accuracy to four decimal places to ensure performance comparable to that of \eqref{TJCs2} and \eqref{TJCs3}. The parameter expressions within \eqref{TJCs2} and \eqref{TJCs3} are delineated as follows:
% \begin{equation*} \label{coef_final}
%     \begin{bmatrix}
%         \bm a_1\\
% 	\bm a_2\\
%         \bm a_3\\
% 	\bm b_1\\
% 	\bm b_2\\
%         \bm b_3        
%     \end{bmatrix}
%     =
%     \begin{bmatrix}
%     -0.004 & 0.027 & 0.056 & 0.934 & 0.000 & -0.060 & 0.988 \\
%     -3.350 & 4.208 & -1.187 & -0.429 & -0.001 & 0.064 & 0.991 \\
%     -0.423 & -3.541 & 6.971 & -6.001 & 2.270 & -121.235 & -0.357 \\
%     -0.018 & 0.058 & 0.023 & 0.389 & 0.000 & -0.032 & 0.989 \\
%     2.628 & -5.189 & 4.072 & -1.436 & -0.099 & 9.002 & 1.086 \\
%     2.228 & -4.659 & 7.736 & -7.937 & 3.328 & -50.958 & 0.421
% \end{bmatrix}
%     \begin{bmatrix}
% 	\alpha_i^3\\
%         \alpha_i^2\\
%         \alpha_i^1\\
%         1\\
%         {\alpha_i}^{-1}\\
%         10^{-5}{\alpha_i}^{-2}\\
%         10^{-10}{\alpha_i}^{-3}
%     \end{bmatrix}.
% \end{equation*}

\section{Conclusions}\label{sec: conclusion}

Based on the kinetic equation, we have investigated the translational and internal temperature profiles in the volumetric heating problem, where strong translational-internal non-equilibrium is present. Then, in the near-continuum flow regime, we have solved the energy equations to obtain the analytical expressions for the temperature jump, with  the second-order temperature-jump boundary conditions. Finally, using the particle swarm optimization algorithm, TJCs have been extracted over a wide range of Knudsen numbers, internal collision number, and accommodation coefficient. These TJCs, given in the analytical form, will be useful in hypersonic flows, where strong translational-internal non-equilibrium are present~\citep{Hypersonic2019,liu2024shock}.

%It exhibits consistent accuracy with the kinetic model for all range of Knudsen number under the Maxwell's specular-diffuse boundary, different translational/internal energy accommodation coefficients and the limit of vanishing internal energy accommodation in near-continuum regime.

\vspace{0.2cm}
\textbf{Acknowledgements:}
This work is supported by the National Natural Science Foundation of China (12172162).

\vspace{0.2cm}
\textbf{Declaration of interest:}
The authors report no conflict of interest.

\bibliographystyle{jfm}
\bibliography{ref}
\end{document}